# A tunable carbon nanotube electromechanical oscillator

Vera Sazonova*, Yuval Yaish*, Hande Üstünel, David Roundy, Tomás A. Arias & Paul L. McEuen

*Laboratory of Atomic and Solid-State Physics, Cornell University, Ithaca, New York 14853, USA*

*These authors contributed equally to this work

**Nanoelectromechanical systems (NEMs) hold promise for a number of scientific and technological applications. In particular, NEMs oscillators have been proposed for use in ultrasensitive mass detection[1,2], radio-frequency signal processing[3,4], and as a model system for exploring quantum phenomena in macroscopic systems[5,6]. Perhaps the ultimate material for these applications is a carbon nanotube. They are the stiffest material known, have low density, ultrasmall cross-sections and can be defect-free. Equally important, a nanotube can act as a transistor[7] and thus may be able to sense its own motion. In spite of this great promise, a room-temperature, self-detecting nanotube oscillator has not been realized, although some progress has been made[8–12]. Here we report the electrical actuation and detection of the guitar-string-like oscillation modes of doubly clamped nanotube oscillators. We show that the resonance frequency can be widely tuned and that the devices can be used to transduce very small forces.**

Figure 1a shows a diagram of the measurement geometry and a scanning electron microscope (SEM) image of a device. The fabrication steps have been described elsewhere[13]; briefly, nanotubes (typically single- or few-walled, 1–4 nm in diameter and grown by chemical vapour deposition[14]) are suspended over a trench (typically 1.2–1.5 μm wide, 500 nm deep) between two metal (Au/Cr) electrodes. A small section of the tube resides on the oxide on both sides of the trench; the adhesion of the nanotube to the oxide[15] provides clamping at the suspension points.

The measurement is done in a vacuum chamber at pressures below $10^{-4}$ torr. We actuate and detect the nanotube motion using the electrostatic interaction with the gate electrode underneath the tube. A gate voltage, $V_g$, induces an additional charge on the nanotube given by $q=C_g V_g$, where $C_g$ is the capacitance between the gate and the tube. The attraction between the charge $q$ and its opposite charge $-q$ on the gate causes an electrostatic force downward on the NT. If $C'_g = dC_g / dz$ is the derivative of the gate capacitance with respect to the distance between the tube and the gate, the total electrostatic force on the tube is given by:

$$F_{el} = \tfrac{1}{2} C'_g V_g^2 \cong \tfrac{1}{2} C'_g V_g^{DC} (V_g^{DC} + 2\delta V_g) \qquad (1)$$

where we have assumed that the gate voltage has both a static (DC) component and a small time-varying (AC) component. The DC voltage $V_g^{DC}$ at the gate produces a static force on the nanotube that can be used to control its tension. The AC voltage $\delta V_g$ produces a periodic electric force, which sets the nanotube into motion. As the driving frequency $\omega$ approaches the resonance frequency $\omega_o$ of the tube, the displacement becomes large.

To detect the vibrational motion of the nanotube, we employ the transistor properties of semiconducting[16] and small-bandgap semiconducting carbon nanotubes[17,18], that is, that the conductance change is proportional to the change in the induced charge $q$ on the tube.

$$\delta q = \delta(C_g V_g) = C_g \delta V_g + V_g \delta C_g \qquad (2)$$

The first term is the standard transistor gating effect—the modulation of conductance due to the modulation of the gate at the driving frequency—and it is observed at any driving frequency. The second term is non-zero only if the tube moves (when the driving frequency approaches the resonance); the distance to the gate changes, resulting in a variation $\delta C_g$ in its capacitance.

To detect this conductance change we use the nanotube as a mixer[19] (Fig. 1b). This method helps avoid unnecessary complications due to capacitive currents between the gate and the drain electrodes. The magnitude of the current is given by the product of the AC voltage on the source electrode $\delta V_{sd}$, and the modulated nanotube conductance $\delta G$. Using equation (2) we derive the result that the expected current is:

$$\delta I^{lock-in} = \delta G \delta V_{sd} = \frac{1}{2\sqrt{2}} \frac{dG}{dV_g} \left( \delta V_g + V_g^{DC} \frac{\delta C_g}{C_g} \right) \delta V_{sd} \qquad (3)$$

where $\delta V_g$ is the AC voltage applied to the gate electrode.

Figure 2a shows the measured current as a function of driving frequency at room temperature. We see a distinctive feature in the current on top of a slowly changing background. We attribute this feature to the resonant motion of the nanotube, modulating the capacitance, while the background is due to modulating gate voltage. The response fits well to a lorentzian function with a normalized linewidth $Q^{-1}=\Delta f/f_o=1/80$, a resonant frequency $f_o$=55 MHz, and an appropriate phase difference between the actuation voltage and the force on the nanotube[19].

The DC voltage on the gate can be used to tune the tension in the nanotube and therefore the oscillation frequency. Figure 2b and c show colour-scale plots of the measured response as a function of the driving frequency and the static gate voltage. The resonant frequency shifts upward as the magnitude of the DC gate voltage is increased. Several distinct resonances are observed, corresponding to different vibrational modes of the nanotube. We have found similar results in 11 comparable devices, with resonance frequencies varying from 3 to 200 MHz for different samples and gate voltages.

To understand the frequency dependence of the nanotube oscillations with $V_g^{DC}$, we have performed a series of simulations of the vibrational properties of nanotubes. We model the nanotube as a slack beam suspended over a trench. Slack here means that the tube is longer than the distance between the contacts, which is a result of the nanotube's curvature before suspension. Slack was observed for almost all imaged devices in an SEM (Fig. 1a), and has also been inferred from atomic force microscope (AFM) force measurements of similar samples[13]. A finite element model is then used to calculate the vibrational frequencies for a nanotube with a typical geometry (length $L$=1.75 μm, radius $r$=1 nm) and mechanical rigidities determined using the Tersoff–Brenner potential[20].

The theoretical results for a representative device can be seen in Fig. 2d. For no static electric force on the nanotube, $V_g^{DC} \approx 0$, the resonance frequency is determined by the bending rigidity of the nanotube and is approximately that of an equivalent doubly clamped beam with no tension. At small $V_g^{DC}$, there is a static electric force downward on the nanotube (equation (1)), producing a tension $T \propto V_g^2$ which shifts the resonant frequency of the nanotube, $\Delta \omega_0 \propto T \propto V_g^2$ (ref. 21). At intermediate $V_g^{DC}$, the electrostatic force overcomes the bending rigidity and the nanotube behaves as a hanging chain; the profile of the tube forms a catenary. In this regime, the resonance frequency is given by $\omega_0 \propto \sqrt{T} \propto V_g$. In the large electrostatic force regime, the nanotube behaves as an elastic string—the extensional rigidity becomes dominant. In this regime, the resonance frequency is given by $\omega_0 \propto \sqrt{T} \propto V_g^{2/3}$ (ref. 21). The transition point from the bending-dominated regime to the catenary, and from the catenary to the stretching-dominated regime, depends on the amount of slack in the nanotube. Either one, two or all three described regimes may be relevant for a particular device. For the bending-dominated and catenary regimes, the voltage dependence of the resonant frequencies scales as the slack to the ¼ th power (Fig. 2d).

Comparing these predictions with Fig. 2b and c, we see a good qualitative agreement with predicted dispersions. All of the resonances start dispersing parabolically, some continuing into a linear regime as the gate voltage is increased. For the lowest resonance shown in Fig. 2c we can also observe the $\omega_0 \propto V_g^{2/3}$ frequency dependence at large gate voltages. The frequency dependence of the resonances are thus in good qualitative agreement with theoretical expectations. We do, however, often find multiple resonances lower in frequency than is predicted by the theoretical calculations. One low-frequency mode is expected because any small asymmetric clamping will result in a non-zero frequency at zero gate voltage for the lowest branch in Fig. 2d. The additional low-frequency modes could be caused by extra mass due to contaminants coating the nanotube, or by a large asymmetry in the clamping conditions. Further studies are needed to understand the exact nature of this frequency lowering.

To determine the other parameters of the nanotube oscillator, we have studied the dependence of the measured resonance on the amplitude of the gate drive signal $\delta V_g$. Figure 3a shows results for one device. For low driving amplitudes, the response on resonance is linear in $\delta V_g$ and $Q$ is roughly constant. As the $\delta V_g$ is increased further, the response saturates and the quality factor decreases. For some devices, there is also a dramatic change in the signal shape observed at these high driving voltages (Fig. 3b). Instead of a smooth lorentzian dip, the system develops a hysteretic transition between low- and high-amplitude states of oscillation.

To understand these results, we first address the linear response regime. We estimate the amplitude $\delta z$ of the nanotube oscillation using the measured signal amplitude relative to the background in conjunction with equation (3). From this, we can extract the relative change in the capacitance $\delta C_g/C_g$ on resonance; for the data in Fig. 2a, where $\delta V_g$=7 mV, we obtain $\delta C_g/C_g$=0.3% . Assuming a logarithmic model for capacitance $C_g = \dfrac{4\pi\varepsilon_0 L}{2\ln(2z/r)}$, where $L$ is the suspended length of the NT, and $z$ is the distance to the gate, this can be translated into a distance change, $\dfrac{\delta z}{z} = \dfrac{\delta C_g}{C_g}\ln(2z/r)$. In Fig. 2a, we estimate the amplitude of motion to be $\delta z \approx 10$ nm. Calculating the driving force using equation (1), we get $F = C'_g V_g^{DC} \delta V_g \approx 60\, fN$. Thus, we estimate the effective spring constant for this resonance to be $k_{eff} = \dfrac{F}{\delta z}Q \approx 4\times 10^{-4}\,\text{N m}^{-1}$. Note that this effective spring constant is different for each resonance.

As the amplitude of the oscillation is increased, we can expect the nonlinear effects due to the change in spring constant to become important. It is well known that nonlinear oscillators have a bistable region in their response-frequency phase space which experimentally results in a hysteretic response[22]. The onset of nonlinear effects in our case corresponds to driving voltages of 15 mV. Assuming the same parameters as above yields an amplitude of motion of 30 nm.

Another important parameter characterizing the oscillator is the quality factor $Q$, the ratio of the energy stored in the oscillator to the energy lost per cycle owing to damping. Maximizing $Q$ is important for most applications. It is in the range of 40–200 for our samples with no observed frequency dependence. Previous measurements on larger multiwalled nanotubes at room temperature and ropes of single-walled nanotubes at low temperatures yielded[8–11] values of $Q$ in the range 150–2,500.

Because one source of dissipation could be air drag, we have studied the dependence of the resonator properties on the pressure in the vacuum chamber. Figure 3c summarizes the results for one device. $Q$ decreases with pressure, and the resonance is no longer observed above pressures of 10 torr. This is in good agreement with calculations[23]. At lower pressures, air losses should be minimal. Many other sources could be contributing to the damping, including the motion of surface adsorbates and ohmic losses due to the motion of electrons on and off the tube. The former is difficult to estimate, but we have calculated the magnitude of the latter and find it to be insignificant. Another important potential source of dissipation is clamping losses where the nanotube is attached to the substrate; the tube may lose energy by sticking and unsticking from the surface during oscillation. Experiments on devices with different clamping geometries are necessary to investigate this issue.

The nanotube oscillator parameters presented above are representative of all of our measured devices. Using these parameters, we can calculate the force sensitivity of the device at room temperature. The smallest detected motion of the nanotube was at a resonant driving voltage of $\delta V_g \approx 1$ mV in the bandwidth of 10Hz. The sensitivity was limited by the Johnson–Nyquist electronic noise from the nanotube. Using equations (1) and (3) above, this corresponds to a motion of ~0.5 nm on resonance and a force sensitivity of ~1 fN Hz$^{-1/2}$. This is within a factor of ten of the highest force sensitivities measured at room temperature[24].

The ultimate limit on force sensitivity is set by the thermal vibrations of the nanotube. The corresponding force sensitivity is $\delta F_{min} = \sqrt{\dfrac{4 k_B kT}{\omega_0 Q}} = 20$ aN Hz$^{-1/2}$ for typical parameters. The observed sensitivity is 50 times lower than this limit. This is probably due to the relatively low values of transconductance for the measured nanotubes at room temperature. At low temperatures (~1 K), the sensitivity should increase by orders of magnitude owing to high transconductance associated with Coulomb oscillations[19]. Even without increasing $Q$, force sensitivities below 5 aN should theoretically be attainable at low temperatures. This is comparable to the highest sensitivities measured[25–28]. The combination of high sensitivity, tunability, and high-frequency operation make nanotube oscillators promising for a variety of scientific and technological applications.

**Acknowledgements** We thank E. Minot for discussions. This work was supported by the NSF through the Cornell Center for Materials Research and the NIRT program, and by the MARCO Focused Research Center on Materials, Structures, and Devices. Sample fabrication was performed at the Cornell Nano-Scale Science and Technology Facility (a member of the National Nanofabrication Infrastructure Network), funded by the NSF.

**Correspondence** and requests for materials should be addressed to L.M. (mceuen@ccmr.cornell.edu).

**Figure 1** Device geometry and diagram of experimental set-up. **a**, A false-colour SEM image of a suspended device (top) and a schematic of device geometry (bottom). Scale bar, 300 nm. Metal electrodes (Au/Cr) are shown in yellow, and the silicon oxide surface in grey. The sides of the trench, typically 1.2–1.5 μm wide and 500 nm deep, are marked with dashed lines. A suspended nanotube can be seen bridging the trench. CVD growth is known to produce predominantly single- and double-walled nanotubes, but we did not perform detailed studies of the number of walls for the nanotubes on our samples. **b**, A diagram of the experimental set-up. A local oscillator(LO) voltage $\delta V_{sd}^{\omega+\Delta\omega}$ (usually around 7 mV) is applied to the source(S) electrode at a frequency offset from the high frequency (HF) gate voltage signal $\delta V_g^{\omega}$ by an intermediate frequency $\Delta\omega$ of 10 kHz. The current from the nanotube is detected by a lock-in amplifier through the drain electrode (D), at $\Delta\omega$, with time constant of 100 ms.

**Figure 2** Measurements of the resonant response. The measurements were done on 11 devices, both semiconducting and small bandgap semiconducting nanotubes in a vacuum chamber at pressures below $10^{-4}$ torr. The maximum conductance $G^{max}$, and the transconductance $dG/dV_g^{max}$, are given below for the presented devices. **a**, Detected current as a function of driving frequency taken at $V_g$=2.2 V, $\delta V_g$=7 mV for device 1 ($G^{max}$=12.5 µS, $dG/dV_g^{max}$=7 µS V$^{-1}$). The solid black line is a lorenzian fit to the data with an appropriate phase shift between the driving voltage and the oscillation of the tube. The fit yields the resonance frequency $f_o$=55 MHz, and quality factor $Q$=80. **b, c**, Detected current (plotted as a derivative in colour scale) as a function of gate voltage and frequency for devices 1 and 2 ($G^{max}$=10 µS, $dG/dV_g^{max}$=0.3 µS V$^{-1}$). Panel **a** is a vertical slice through panel **b** at $V_g$=2.2 V (marked with a dashed black line). The insets to the figures show the extracted positions of the peaks in the frequency–gate voltage space for the respective colour plots. A parabolic and a $V_g^{2/3}$ fit of the peak position are shown in red and green, respectively. **d**, Theoretical predictions for the dependence of vibration frequency on gate voltage for a typical device with length $L$=1.75 µm, and radius $r$=1 nm. The calculations were performed for several different values of slack $s$ ($s$=($L-W$)/$W$, where $L$ is the tube's length and $W$ is the distance between clamping points). The calculations for 0.5%, 1% and 2% slack are shown in blue, red and green, respectively. Notice the appropriately rescaled $x$-axis.

**Figure 3** Amplitude and pressure dependence of the resonance. **a** The measured quality factor $Q$ of the resonance and the height of the resonance peak for device 3 ($G^{max}$=15 µS, $dG/dV_g^{max}$=4 µS V$^{-1}$) are shown in red open squares and black solid squares, respectively, as a function of driving voltage $\delta V_g^\omega$. Linear behaviour is observed at low voltages, but $Q$ decreases and the height of the peak saturates at higher driving voltages. **b**, Trace of detected current versus frequency with the background signal subtracted for device 2 at two different driving voltages $\delta V_g$=8.8 mV and $\delta V_g$=40 mV. The solid black line is a lorenzian fit to the low bias data. The traces of the current as the frequency is swept up and down are shown in blue and black, respectively. Hysteretic switching can be observed. **c**, Pressure dependence of the resonance peak for device 4 ($G^{max}$=7.7 µS, $dG/dV_g^{max}$=0.6 µS V$^{-1}$). The $Q$ of the resonance peak is shown in red open squares. The peak was no longer observed above pressures of 10 torr.

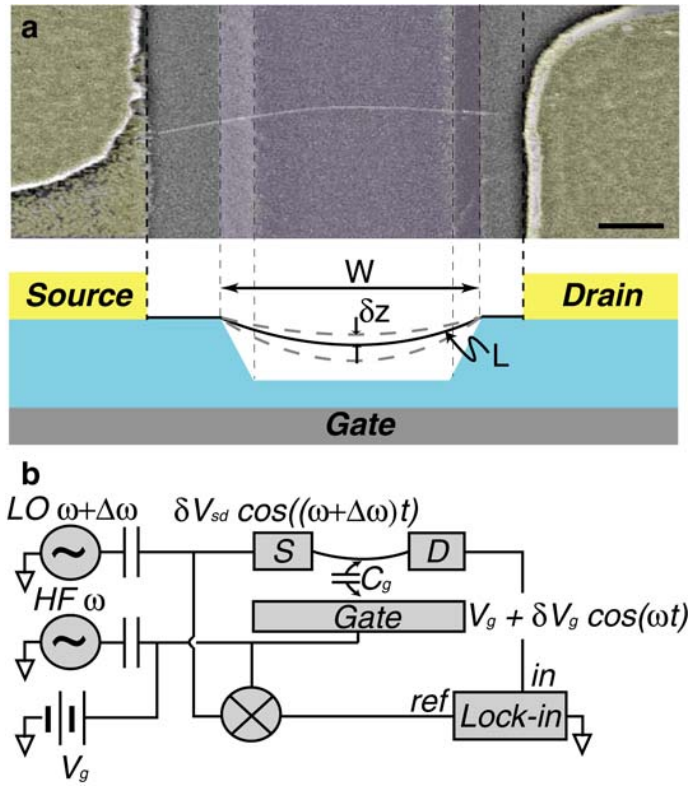

Figure 1.

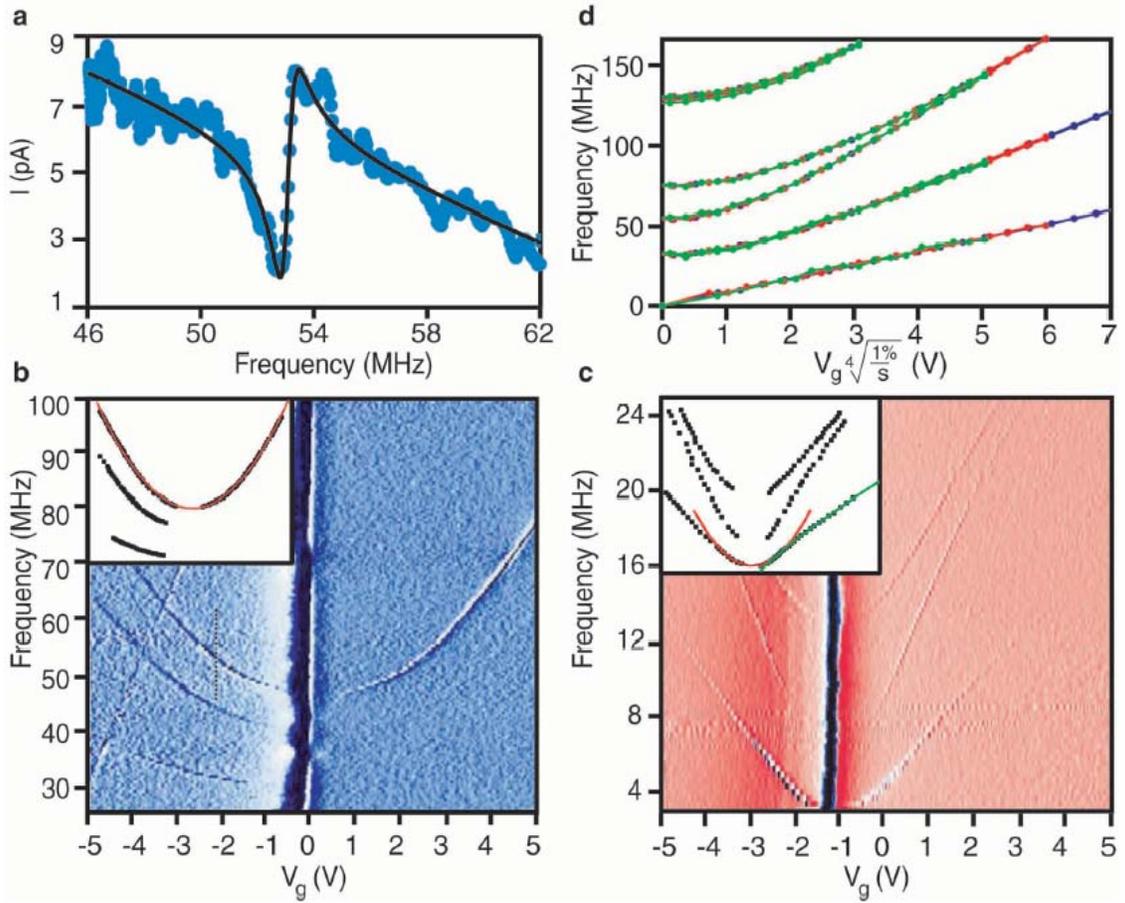

Figure 2.

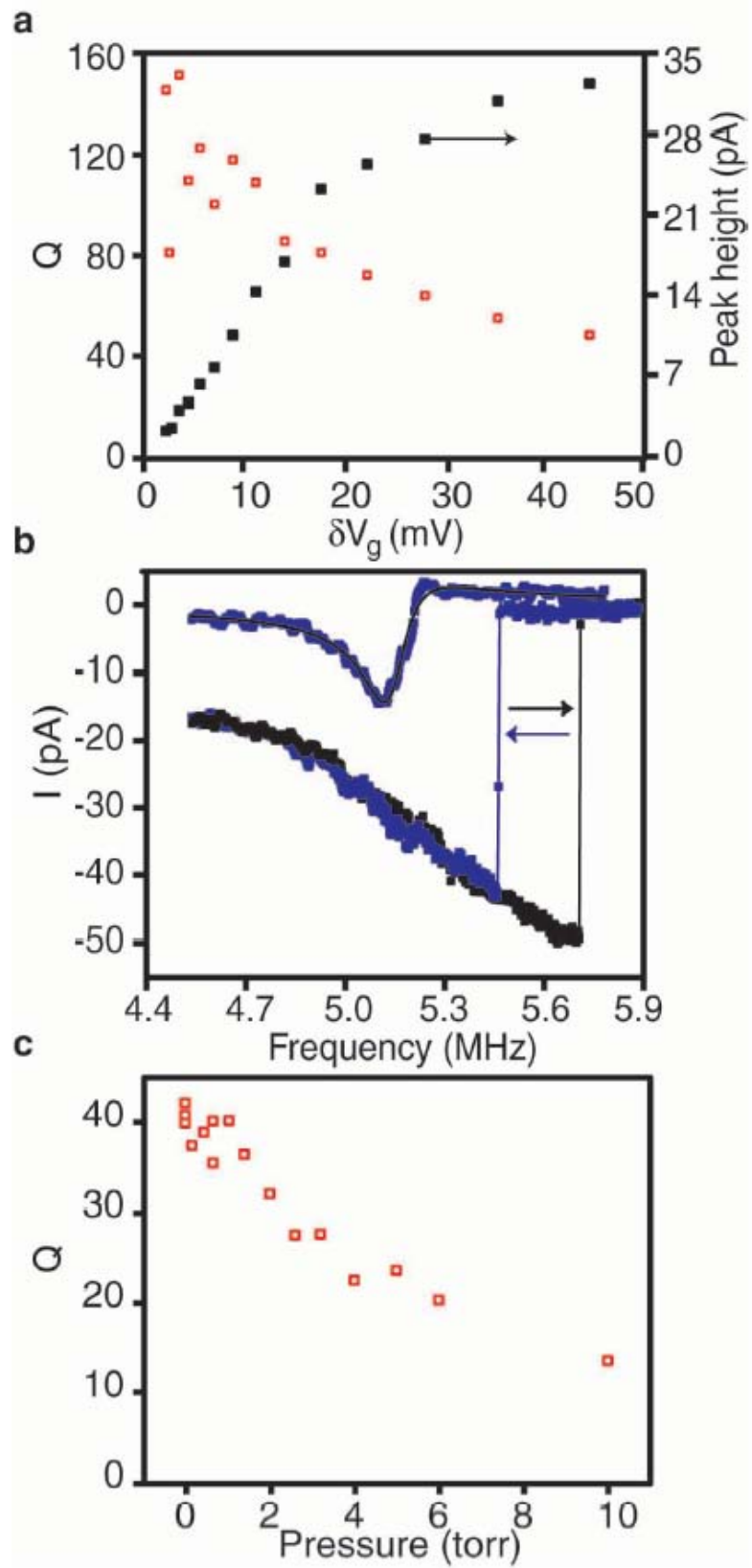

Figure 3.